\documentstyle[11pt,epsfig]{article} 

\def\n55{n_{5\overline 5}}
\def\tb{\tan\beta}
\def\gy{\tilde g_1}
\def\gl{\tilde g_2}
\def\gc{\tilde g_3}
\def\l{\tilde \lambda}
\def\k{\tilde k}
\def\htt{\tilde h_t}
\def\hb{\tilde h_b}
\def\htau{\tilde h_\tau}
\def\QQa{\renewcommand{\baselinestretch}{1.3}\Huge\large\normalsize}

\begin{document}

\pagestyle{empty}
\begin{obeylines}
\begin{flushright}
UG-FT-84/97\\
UAB-FT-436\\
January 1998
\end{flushright}
\end{obeylines}
\vspace{2cm}

\begin{center}
\begin{bf}
\centerline {\large \bf Limits on the mass of the lightest Higgs}
\centerline {\large  \bf in supersymmetric models}
\end{bf}
\vspace{1.cm}

M. Masip$^{(1)}$, R. Mu{\~n}oz-Tapia$^{(1)}$ and A. Pomarol$^{(2)}$

\vspace{0.5cm}
{\it $^{(1)}$Departamento de F{\'\i}sica Te{\'o}rica y del Cosmos,
Universidad de Granada,
18071 Granada, Spain\\}

\vspace{0.1cm}
{\it $^{(2)}$Institut de F{\'\i}sica d'Altes Energies, 
Universitat Aut{\`o}noma de Barcelona,
08193 Bellaterra (Barcelona), Spain}
\vspace{1.8cm}

{\bf Abstract}

\parbox[t]{12cm}{
In supersymmetric models extended with a gauge singlet the 
mass of the lightest Higgs boson has contributions proportional
to the adimensional coupling $\lambda$. In minimal scenarios,
the requirement that this coupling remains perturbative up to the
unification scale constrains $\lambda$ to be smaller than 
$\approx 0.7$. We study the maximum value of $\lambda$ 
consistent with a perturbative unification of the gauge couplings 
in models containing nonstandard fields at intermediate scales. 
These fields appear in scenarios with gauge mediation of 
supersymmetry breaking. We find that the presence of extra 
fields can raise the maximum 
value of $\lambda$ up to a 19\%, increasing 
the limits on the mass of the lightest Higgs from 135 GeV 
to 155 GeV.
 }
\end{center}

\newpage
\pagestyle{plain}
\QQa

The main motivation of 
supersymmetric (SUSY) extensions of the standard model
is their stability against quantum corrections. SUSY
models provide a 
framework to integrate large energy scales 
together with the observed 
low-energy physics. This generic motivation
has been recently underlined by the celebrated perturbative
unification of the three gauge couplings 
in the minimal extension (MSSM). 
Up to now, however, there is no observation
in disagreement with the standard model predictions.
Supersymmetry, although attractive from a theoretical
point of view, is still lacking 
experimental confirmation. 

SUSY models
have been {\it flexible} enough to respect all experimental 
constraints, but this flexibility does not translate into
a complete lack of low-energy predictivity. 
The most compelling prediction of SUSY models is probably
the presence of a light Higgs field. In particular, the
MSSM  forces the $CP$-even scalar field $h^0$ 
to have a tree-level mass $m_h$ smaller than $M_Z$:
\begin{equation}
m_h^2 \le M_Z^2\;\cos^22\beta\;, 
\label{mh0}
\end{equation}
where $\tan\beta$ is the ratio of vacuum expectation values
(VEVs) $v$ and $\bar v$ of the Higgs fields $H$ and $\bar H$
that give mass to the up and down type quarks, 
respectively (see \cite{hhg90} for a review). 
This tree-level bound is shared
by any SUSY model with only doublets in the Higgs sector
\cite{cha85}. 

In models with gauge singlets, trilinear terms in the
superpotential of the type 
\begin{equation}
W \supset \lambda\; SH\bar H 
\label{p1}
\end{equation}
introduce 
new quartic interactions for the scalar Higgs doublets. 
The tree-level bound becomes
\begin{equation}
m_h^2 \le M_Z^2\;\cos^22\beta + \lambda^2 \nu^2\;\sin^22\beta \;, 
\label{mh1}
\end{equation}
with $\nu=\sqrt{v^2+\bar v^2}=174$ GeV.
The impact of this new term, however, is 
limited by the following argument \cite{esqui}. The $\beta$-function
fixing the running of $\lambda$ is at one loop
\begin{equation}
\beta_\lambda = {\lambda\over 16 \pi^2} (
4 \lambda^2 + 3 h_t^2 + 3 h_b^2 - g_1^2 - 3 g_2^2)\;, 
\label{beta}
\end{equation}
where $h_t$ and $h_b$ are the top and bottom Yukawa couplings,
and $g_1$ and $g_2$ are the $U(1)_Y$ and $SU(2)_L$ gauge
couplings, respectively. The evolution of $\lambda$ 
will be dominated by $h_t$, which means
that its value increases with the energy. 
As a  consequence, the value of $\lambda$ at the weak scale must 
be small if we want to be 
in the perturbative regime 
up to the grand unification scale $M_X=1.4\times 10^{16}$ GeV. 
For the top quark observed at
CDF \cite{pdb96}, this argument implies that the
low-energy value of $\lambda$ must be smaller
than $\approx 0.7$ \cite{kin95}. Moreover, 
any Yukawa coupling that can be added to the superpotential, 
like trilinears 
\begin{equation}
W \supset -{1\over 3} k \; S^3\;,
\label{p2}
\end{equation}
gives a  positive contribution to 
$\beta_\lambda$ and further decreases the maximum value of $\lambda$
that remains perturbative up to $M_X$. 

A possible way  to increase the value of $\lambda$  
(and consequently $m_h$)
is to introduce new matter fields at
intermediate scales. The effect of these fields
on $\lambda$ would be indirect, in the
sense that they increase the evolution rate
of the gauge couplings, which in turn
decreases the evolution rate of $\lambda$. 
Note that $g^2_1$ and $g^2_2$ contribute negatively to 
$\beta_\lambda$; larger values of these couplings imply
a slower running of $\lambda$ and then that larger initial
values of this coupling would remain perturbative up to $M_X$.
This argument was outlined by Kane and collaborators in 
Ref.~\cite{kan93}. They only introduced extra 
Higgs doublets because only $g^2_1$ and $g^2_2$ (and not $g^2_3$)
appear in $\beta_\lambda$. 
They found that the effect on $\lambda$
is always small. A sizeable effect would require
the inclusion of many doublets at low-energy scales, but then
the gauge couplings become non-perturbative before $M_X$.
In addition, the presence of Higgs doublets spoils
the unification of the gauge couplings observed in
the MSSM, which constitutes so far the only phenomenological 
motivation for supersymmetry.
 
The previous analysis, nevertheless, can be improved. First,
it will be convenient to introduce extra quarks as well as 
extra leptons (or Higgs fields). 
The addition of matter with $SU(3)_C$
interactions will raise the intermediate values of $g_3$. 
Since the contribution of 
$g_3$ enters in $\beta_{h_t}$ with
negative sign, 
larger values of $g_3$ will imply lower values of $h_t$, 
which would decrease the evolution rate of $\lambda$ (note that 
$h_t\approx g_3\approx 1$ are the 
dominant couplings in the renormalization group equations). 
Second, 
there is a way to introduce extra matter fields that respects
the perturbative unification of the three gauge couplings
of the MSSM. If we add complete representations of a simple
group [$SU(5)$, $SO(10)$, $E_6$,...] that contains
 $SU(3)_C\times SU(2)_L\times U(1)_Y$ 
as a subgroup, then the 
(one-loop) effect 
on the running of the three gauge couplings is such that 
the couplings still meet at the same unification scale $M_X$
but with a higher final value. This fact, well known by the 
practitioners, allows larger intermediate values of $g_1$ and
$g_2$ together with smaller values of $h_t$ (via larger $g_3$),
and will define the scenario for the absolute 
perturbative bound on $\lambda$. In Fig.~1 we plot the
running of the gauge couplings in the MSSM and in a model
extended with four families of $5+\overline 5$ at 1 TeV.

The presence of matter in vectorlike representations of the
standard model symmetry finds its primary motivation on 
models with gauge mediated supersymmetry breaking (GMSB)
\cite{review}. 
The extra fields, called {\it messengers} $\Phi$-$\bar\Phi$, 
have a  mass $M$ that can  vary from $\sim 30$ TeV
to $M_X$, and couple directly to
the fields that break supersymmetry (to the {\it secluded} sector).
This coupling induces a scalar-fermion mass splitting $\sqrt{F}$
inside the messenger superfields that is 
transmitted, at the loop level, to the standard superfields.
Actually, the minimal  
scenarios for GMSB could be closer to the singlet model than
to the MSSM. The reason is that 
these models have serious difficulties  
to generate the $\mu$ 
term (the Higgsino mass) in the superpotential \cite{dgp},
and usually require the presence of non-standard 
fields and couplings.
One simple possibility \cite{dn} is
to introduce a singlet superfield with the coupling in 
Eq.~(\ref{p1}) and generate 
$\mu=\lambda\langle S\rangle$ via VEVs.
We will discuss later in some detail aspects of the singlet model
which are specific to GMSB scenarios.

In this rapid communication 
we study the bounds on $\lambda$ in scenarios with vectorlike
fields at intermediate scales. 
The couplings involved in our analysis are $g_1$, $g_2$, $g_3$, 
$h_t$, $h_b$, $h_\tau$, and $\lambda$.
We take a top quark mass of 180 GeV (pole mass) and 
$\alpha_s(M_Z) = 0.118$ \cite{pdb96}. The extra matter present at
a given scale 
is parametrized by the number $\n55$ of 
$5+\overline 5$ representations of $SU(5)$, which is the lowest
dimensional vector representation of a simple group containing
the standard model symmetry.
In the appendix we include the 
two-loop renormalization group equations for these parameters.
For different values
of $\tb$, we will look for the highest 
low-energy value of $\lambda$ consistent with a perturbative
value of all the parameters up to the unification scale.
Note that $\beta_\lambda$ 
includes a negative two-loop contribution $-10\lambda^4$.
In consequence, at this order $\lambda$ does not have the 
ultraviolet Landau pole of 
non-SUSY $\phi^4$ theories. Here the large 
values of $\lambda$ will grow with the energy scale 
but only up to the point where the 
one-loop and the two-loop contributions to $\beta_\lambda$ cancel. 
This value will correspond 
approximately to ${\lambda\over 4\pi}=\sqrt{4/10}$.
We shall then consider that $\lambda$ 
is non-perturbative if at a scale below $M_X$ the running value 
is ${\lambda\over 4\pi} > 0.3$. The same
criterion will be used for the $h_t$ and $h_b$. The change from the
perturbative to the non-perturbative regime is quite abrupt, and 
therefore the results do not depend on the actual maximum value for
the running couplings at $M_X$ that we choose.

Let us start analyzing the minimal
model with no extra matter at intermediate scales. 
To obtain the bound on $\lambda$ we take $k=0$. 
The lowest allowed value of $\tb$ is 1.88, as for
smaller values $h_t$ becomes non-perturbative [i.e.,
${h_t(M_X)\over 4\pi}>0.3$] even if $\lambda=0$. For larger
values of $\tb$ the initial value of $\lambda$
is constrained by the simultaneous conditions that $h_t$ and
$\lambda$ remain perturbative up to $M_X$. For $\tb$ up to 2.31
the dominant condition is that $h_t$ remains perturbative, whereas
for $(2.31<\tb<59.81)$ $\lambda$ itself is the first coupling
to become non-perturbative. For very large values of $\tb$,
from 59.81 to 61.23, the dominant condition is that $h_b$ remains 
perturbative up to $M_X$. We plot in Fig.~(2) the maximum value 
of $\lambda$ for each $\tan\beta$. The absolute limit is
$\lambda<0.69$, which corresponds to $\tb=10$. 

When extra matter is included the intermediate values of the
gauge couplings grow [see Fig.~(1)] decreasing $\beta_\lambda$. 
To find the absolute limit on $\lambda$, 
we add the maximum number $\n55$ of
$5+\overline 5$ families
consistent with a perturbative unification of
the gauge couplings. 
It turns out that we can add four families at 250 GeV, or 
four families at one TeV plus another one at $10^{11}$ GeV,
or five families at 60 TeV. We plot in 
Fig.~(2) the first case, although the limits are similar 
in the other two cases. 
The maximum value $\lambda=0.82$, obtained for $\tb= 8$,
is a 19\% higher than in the case with no vectorlike matter at
intermediate scales.
We also observe that lower values of $\tb$ are 
possible without going to non-perturbative values of $h_t$;
here the limit is $\tb=1.19$ versus $\tb=1.88$ in the MSSM. 
This fact
is remarkable because the main contribution to $m_h$ in 
the singlet model comes at low $\tb$.

It is now straightforward to
translate this maximum values of $\lambda$
into the limits on the mass of the lightest Higgs boson. In addition
to the tree-level contributions in Eq.~(3), we 
include 
top-quark radiative corrections. Following
the procedure described in \cite{kin95}, one obtains that 
radiative corrections contributing to $m_h^2$ vary from
(95 GeV)$^2$ at $\tb=1.2$ to (90 GeV)$^2$ at $\tb=74$.
In Fig.~(3) we plot the bound to $m_h$ in the MSSM, in the
singlet model with no matter, and in the singlet model with a
maximum content of vectorlike matter. As it is 
apparent from this figure,
the bounds on $m_h$ are considerably relaxed;
if the presence of a singlet takes the MSSM bound
from $m_h\le 128$ GeV to 
$m_h\le 135$ GeV,
the presence of matter at intermediate scales pushes 
this bound further up, to $m_h\le 155$ GeV. 

Since one of the motivations for enlarging
the MSSM with the singlet $S$ and 
extra vectorlike fields arises from GMSB models,
we would like to make a final remark on
the viability of these theories. In minimal scenarios of GMSB 
one has that the trilinear soft masses are
smaller than the other soft masses \cite{review}. As a consequence, 
these models suffer from the presence of 
a too light scalar field \cite{dn,gfm}. 
A possible solution to this problem is 
to introduce mixing between the 
messenger and the ordinary matter sectors \cite{cp,gr}.
Then the scalar trilinears are 
induced at the one-loop level, as the other soft masses. 
There are different possibilities. One could introduce
a messenger-matter mixing from the couplings 
$HQ\bar\Phi$, $S\Phi\bar\Phi$, $H\Phi\bar\Phi$ or
$SH\bar\Phi$ (and equivalently for
$H\rightarrow \bar H$ ) \cite{dgp,gr,dns}. 
Any of them induces at one loop scalar trilinears.
For example, the coupling
$W \supset \lambda'\; H\Phi\bar\Phi$ gives
to the trilinear and scalar mass the new contributions
\begin{eqnarray}
\delta A&=&-\frac{\lambda^{'2}}{16\pi^2}\frac{F}{M}\label{trimass}\,\\
\delta m^2_H&=&\frac{-\lambda^{'2}}{48\pi^2}\frac{F^4}{M^6}+
\frac{\lambda^{'2}}{256\pi^4}(4\lambda^{'2}+3h_t^2-
\frac{3}{5}g_1^2-3g_2^2)\frac{F^2}{M^2}\, ,
\label{softmass}
\end{eqnarray}
where we are assuming that 
$F$ and $M$ are the same for the two messenger fields coupled
to $H$ and $F<M^2$ (if this is not the case, see ref.~\cite{dgp}).
The first term of eq.~(\ref{softmass}) enters at the one-loop level
but it is suppressed for $F<M^2$;
the second term, calculated in 
Ref.~\cite{gr}, arises at the 
two-loop level.
We have checked that, after including the above soft mass 
contributions, there are regions of the
parameter space of the model 
that lead to phenomenologically viable scenarios of electroweak 
symmetry breaking.
In particular, we find that the model
proposed in ref.~\cite{cp}, with
$S$  playing the role of sliding singlet \cite{witten}, 
has a (at least  local) minimum in which $\mu$ is of the order 
of the weak scale\footnote{It is not clear if this model could 
survive after the recent LEP2 bounds on the superpartners masses.}. 
Of course, a certain degree of fine tuning
of the parameters is required in order to obtain an acceptable minimum, 
but this is a generic problem of GMSB theories.
Different conclusions were reached in 
ref.~\cite{gfm}, but there 
the  contribution (\ref{softmass}) to the scalar soft masses
was not included.

From the above we conclude that 
GMSB scenarios with a singlet 
require extra couplings of the Higgs with 
the intermediate matter (messengers). 
These couplings will modify the 
running of $\lambda$ discussed previously. 
For example, 
the inclusion of a new leptonic coupling $\lambda'=0.5$ at 30 TeV
will move the maximum low-energy value of $\lambda$ a 2\% down.
However, the presence of extra matter will 
increase the maximum values significantly also in this
case. In particular, 
adding four $5+\overline 5$ families at the same 
energy scale would allow values  of $\lambda$ 
a $10\%$ larger, up to  $\lambda=0.74$. 

In summary, 
we have analyzed how the presence of vectorlike fields
relaxes the bounds
on $m_h$ in extensions of the MSSM
with a gauge singlet. These scenarios could be
motivated by GMSB. The gauge
unification of the minimal model is preserved and the range
of $\tb$ consistent with a perturbative value of $h_t$ and
$h_b$ up to $M_X$ is extended. The mass of the lightest
Higgs can be raised from 135 to 155 GeV. This is 
a correction that does not change the generic feature of 
SUSY models, the presence of a light neutral Higgs. However,
since this corresponds to the 
 range of energies to be explored in the near
future, its phenomenological impact can be important.

\section*{Acknowledgments} 

We thank F. del Aguila for helpful comments. 
The work of M.M. and R.M.T.  
was supported by CICYT under contract 
AEN96-1672 and by the Junta de Andaluc{\'\i}a under contract
FQM-101. The work of A.P. was supported by CICYT under contract 
AEN95-0882.

\section*{APPENDIX} 

In this appendix we write the two-loop renormalization group
equations for the couplings involved in our analysis.
The parameter $\n55$ expresses
the number of $5+\overline 5$ representations of $SU(5)$ 
present at a given scale. The equations have been deduced
from the general expressions given in \cite{mar94}. To obtain
the initial values of $h_t$ and $h_b$ from the pole masses,
see \cite{agu96}. We denote with a tilde the parameter over
$4\pi$: $\tilde g=g/4\pi$, 
and $t=\ln \mu$.
For the three gauge couplings we have
\begin{eqnarray}
{d\over dt}\gy &=& ({33\over 5}+\n55)\;\gy^3+\gy^3\;[\;
({199\over 25}+{7\over 15}\n55)\;\gy^2+
({27\over 5}+{6\over 5}\n55)\;\gl^2+
({88\over 5}+{32\over 15}\n55)\;\gc^2\nonumber \\
&&-{26\over 3}\htt^2-{14\over 3}\hb^2
-{6}\htau^2-{2}\l^2\;]\nonumber \\
{d\over dt}\gl &=& (1+\n55)\;\gl^3+\gl^3\;[\;
({9\over 5}+{3\over 5}\n55)\;\gy^2+
({25}+{3}\n55)\;\gl^2 + 24\gc^2 
-6\htt^2\nonumber \\
&&-6\hb^2-2\htau^2-2\l^2\;]\nonumber \\
{d\over dt}\gc &=& (-3+\n55)\;\gc^3+\gc^3\;[\;
({11\over 5}+{8\over 5}\n55)\;\gy^2+
9\gl^2+(14+{16\over 3}\n55)\;\gc^2
-4\htt^2-4\hb^2\;]\;.\nonumber \\
\label{rge1}
\end{eqnarray}
The equations for $h_t$, $h_b$, $h_\tau$, $\lambda$ and $k$ are
\begin{eqnarray}
{d\over dt}\htt &=& \htt\;(6\htt^2+\hb^2+\l^2-{13\over 15}\gy^2
-3\gl^2-{16\over 3}\gc^2)+\htt\;[\;{2743\over 450}\gy^4+
{15\over 2}\gl^4-{16\over 9}\gc^4\nonumber \\
&&+\gy^2\gl^2+{136\over 45}\gy^2\gc^2+8\gl^2\gc^2+
\htt^2\;({6\over 5}\gy^2+6\gl^2+16\gc^2)+
{2\over 5}\gy^2\hb^2-22\htt^2-\nonumber \\
&&5\htt^2\hb^2-5\hb^4-\hb^2\htau^2-\l^2\;(3\htt^2+4\hb^2+
\htau^2+2\k^2+3\l^2)\;]\nonumber \\
{d\over dt}\hb &=& \hb\;(6\hb^2+\htt^2+\htau^2+\l^2-
{7\over 15}\gy^2-3\gl^2-{16\over 3}\gc^2)+\hb\;[\;
{287\over 50}\gy^4+{15\over 2}\gl^4-\nonumber \\
&&{16\over 9}\gc^4+\gy^2\gl^2+{8\over 9}\gy^2\gc^2+8\gl^2\gc^2+
\hb^2\;({2\over 5}\gy^2+6\gl^2+16\gc^2)+
{4\over 5}\gy^2\htt^2+{6\over 5}\gy^2\htau^2\nonumber \\
&&-22\hb^2-5\htt^2\hb^2-5\htt^4-3\hb^2\htau^2-3\htau^4
-\l^2\;(3\hb^2+4\htt^2+2\k^2+3\l^2)\;]\nonumber \\
{d\over dt}\htau &=& \htau\;(4\htau^2+3\hb^2+\l^2-
{9\over 5}\gy^2-3\gl^2)+\htau\;[\;{243\over 50}\gy^4+
{15\over 2}\gl^4+{9\over 5}\gy^2\gl^2+\nonumber \\
&&\htau^2\;({6\over 5}\gy^2+6\gl^2)+\hb^2\;(-{2\over 5}\gy^2+
16\gc^2\;)\;-10\htau^4-9\hb^2\htau^2-9\hb^4-3\htt^2\hb^2\nonumber \\
&&-\l^2\;(3\htau^2+3\htt^2+2\k^2+3\l^2
)\;]\nonumber \\
{d\over dt}\k &=& \k\;(6\l^2+6\k^2)+\k\;[\;
{18\over 5}\gy^2\l^2+18\gl^2\l^2-\l^2\;(12\l^2-18\htt^2-
18\hb^2-6\htau^2-\nonumber \\
&&24\k^2)-24\k^4\;]\nonumber \\
{d\over dt}\l &=& \l\;(3\htt^2+3\hb^2+\htau^2+4\l^2+2\k^2
-{3\over 5}\gy^2-3\gl^2)+\l\;[\;
{207\over 50}\gy^4+{15\over 2}\gl^4+\nonumber \\
&&{9\over 5}\gy^2\gl^2+{6\over 5}\gy^2\l^2+
{6\over 5}\gy^2\htau^2+{4\over 5}\gy^2\htt^2-
{2\over 5}\gy^2\hb^2+6\gl^2\l^2+
16\gc^2\htt^2+16\gc^2\hb^2-\nonumber \\
&&9\htt^4-6\htt^2\hb^2-9\hb^4-3\htau^4-10\l^4-
8\k^4-\l^2\;(9\htt^2+9\hb^2+3\htau^2+12\k^2
)\;]\nonumber \\
\label{rge2}
\end{eqnarray}

\newpage
\setlength{\unitlength}{1cm}
\begin{figure}[htb]
\begin{picture}(10,12)
\epsfxsize=20cm
\put(-3,-14.){\epsfbox{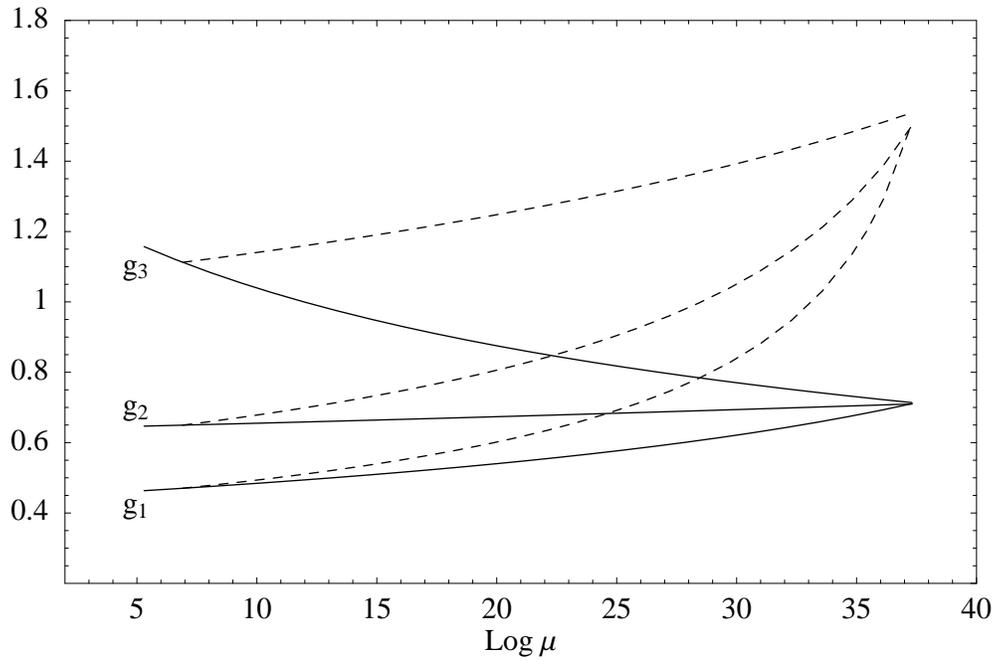}}
\end{picture}
\caption{Evolution of the three gauge couplings in the MSSM (solid) 
and in the model extended with four complete $5+\overline 5$ 
representations of $SU(5)$ at 1 TeV (dashes). The scale $\mu$ is given 
in GeV units. \label{fig1}}
\end{figure}

\newpage
\begin{figure}[htb]
\begin{picture}(10,12)
\epsfxsize=20cm
\put(-3,-14.){\epsfbox{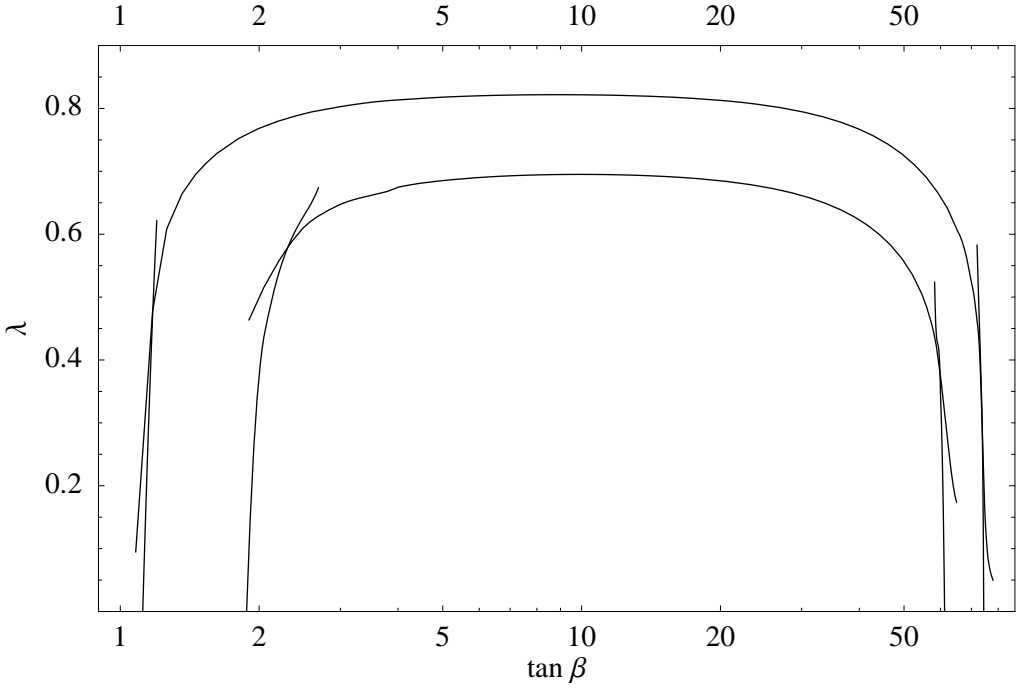}}
\end{picture}
\caption{Limits on the value of $\lambda$ at the weak scale.
We plot the singlet model in the cases with
a maximal matter content  at intermediate scales
(upper) and without extra matter (lower).
\label{fig2}}
\end{figure}

\newpage
\begin{figure}[htb]
\begin{picture}(10,12)
\epsfxsize=20cm
\put(-3,-14.){\epsfbox{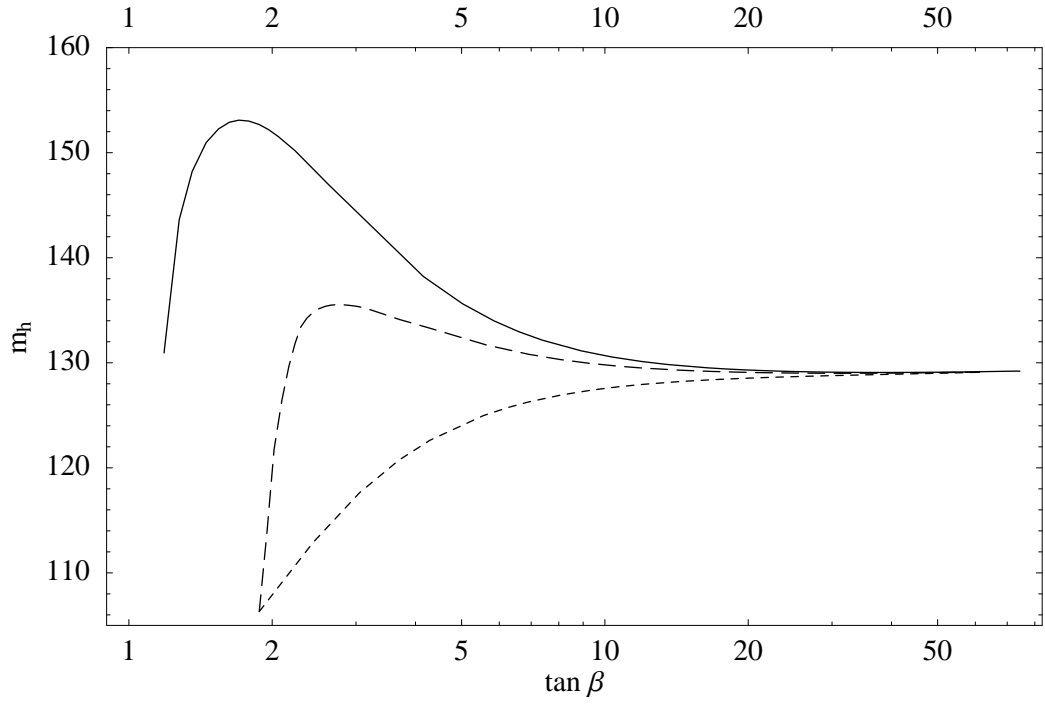}}
\end{picture}
\caption{Limit on $m_h$ (in GeV) in the MSSM, in the singlet model with
no intermediate matter, and in the singlet model with a maximal
matter content at intermediate scales. We have included top
radiative corrections with $(m_{\tilde t}^2+
m_{\tilde t^c}^2)/2=
1$ TeV$^2$. 
\label{fig3}}
\end{figure}

\end{document}